\documentclass[twocolumn,preprintnumbers,amsmath,amssymb,groupedaddress,superscriptaddress,prx]{revtex4}
\usepackage{amsmath,amsfonts,amssymb,graphicx,color,times,bbm}
\usepackage{graphicx}
\usepackage{dcolumn}
\usepackage{bm}
\usepackage{amssymb}
\usepackage{amsmath}
\usepackage[colorlinks=true,citecolor=blue,urlcolor=blue]{hyperref}
\newcounter{fnnumber}
\usepackage{footmisc}
\renewcommand{\emph}{\textit}
\begin{document}

\title{Optimal control for one-qubit quantum sensing}

\author{F. Poggiali}
\affiliation{
LENS European Laboratory for Non linear Spectroscopy, Universit\`a di Firenze, I-50019 Sesto Fiorentino, Italy} \affiliation{
INO-CNR Istituto Nazionale di Ottica del CNR, I-50019 Sesto Fiorentino, Italy}
\author{P. Cappellaro}
\affiliation{
LENS European Laboratory for Non linear Spectroscopy, Universit\`a di Firenze, I-50019 Sesto Fiorentino, Italy}
\affiliation{Department of Nuclear Science and Engineering, Massachusetts Institute of Technology, Cambridge, MA 02139}  
\author{N. Fabbri}\email{fabbri@lens.unifi.it}
\affiliation{
LENS European Laboratory for Non linear Spectroscopy, Universit\`a di Firenze, I-50019 Sesto Fiorentino, Italy}
\affiliation{
INO-CNR Istituto Nazionale di Ottica del CNR, I-50019 Sesto Fiorentino, Italy}

\begin{abstract}
Quantum systems can be exquisite sensors thanks to their sensitivity to  external perturbations. This same characteristic also makes them fragile to external noise. Quantum control can tackle the challenge of protecting  quantum sensors from environmental noise, while leaving their strong coupling to the target field to be measured. As the compromise between these two conflicting requirements  does not always have an intuitive solution,  optimal  control based on numerical search could prove very effective. 
Here we adapt optimal control theory  to the quantum sensing scenario, by introducing a cost function that, unlike the usual fidelity of operation, correctly takes into account  both the unknown field to be measured and the environmental noise. 
We experimentally implement this novel control paradigm using a Nitrogen Vacancy center in diamond, finding improved sensitivity to a broad set of time varying fields. The demonstrated robustness and efficiency of the numerical optimization, as well as the sensitivity advantaged it bestows, will prove beneficial to many quantum sensing applications.

\end{abstract}

\maketitle
\section{Introduction}
Quantum control has been demonstrated to be a crucial tool both in quantum information processing~\cite{Nielsen00b} and in quantum sensing~\cite{Giovannetti11, Degen17} on a variety of experimental platforms, ranging from trapped ions~\cite{Wineland05,Timoney08}, to ultracold atoms~\cite{Jessen01,Rosi13}, superconducting qubits~\cite{Reed12,Sporl07}, as well as nuclear~\cite{Cory00,Zhang11a} and electronic spin qubits~\cite{Wrachtrup06,Waldherr14}. Quantum sensing poses  peculiar challenges to control, as sensor qubits need to interact  strongly  with the target field to be probed, but this also leads to undesired coupling with external noise of the same nature of the target field, which often gives rise to either energy losses or decoherence. A paradigmatic scenario is when one wants to measure a frequency shift of a spin qubit sensor, as due to a magnetic field, in the presence of magnetic dephasing noise. 

Optimal control theory~\cite{Dalessandro07,Glaser15} exploits numerical optimization methods~\cite{Fortunato02,Khaneja05,Caneva11,Machnes11, Ciaramella15}, to find the best control fields that steer the dynamics of a system towards the desired goal. Quantum optimal control has been successfully applied in the case of one- and few-body systems~\cite{Ryan10,Scheuer14,Tannor85, Brumer92, Weinacht99, Machnes10, Rahmani13}, as well as ensembles~\cite{Tosner09} and correlated many-body quantum systems~\cite{Rosi13,Doria11,vanFrank16}. 

Typically, the optimal control problem involves the search for the optimal transformation that, given a system Hamiltonian $\mathcal{H}$ dependent
 on a set of time-dependent control fields, drives the system from an initial state into a target state, whose desired properties are expressed by a cost function $\mathcal{F}$ that one wants to minimize. Often this means maximizing the fidelity of the  unitary operation, which describes this transformation, with the desired one. 

The goal of quantum sensing is however different. Since in principle we do not have any {\it a priori} knowledge of the external field to be measured, we do not know what is the expected unitary dynamics, and thus we cannot use the fidelity to optimize control. In addition, quantum sensing is usually concerned with optimizing sensitivity, a quantity that intrinsically includes noise, also  arising from the external environment. 

Here we devise and experimentally demonstrate a robust and efficient scheme for optimal control of a sensing qubit, which enhances its sensitivity as a probe of time-varying target fields. To this purpose, we use an unconventional optimization metric, the sensitivity, and develop a practical way of computing it (which allows for fast numerical searches). Furthermore our search method does include in the cost metric itself the presence of an environment and consequent decoherence induced on the qubit. While optimal control has been used before for sensing~\cite{Haberle13,Nobauer15}, the optimization was only targeted at improving the control fidelity and bandwidth, not the sensitivity itself.

We tackle here the complex task of measuring multi-chromatic AC target fields, and different significant waveforms, such as trains of magnetic impulses, which are relevant for applications in biology, physiology, and neuroscience~\cite{Wikswo80,Bison09,Jensen16,Barry16}. We show that in these cases optimal control demonstrates better performance than traditional dynamical decoupling, since it allows for both a larger accumulation of the spin phase that encodes the field information, and for an improved compensation of environment-induced decoherence, thus boosting qubit's sensitivity and enabling detection of very weak magnetic fields.

\begin{figure*}[t!]
\begin{center}
\includegraphics[width=.7\textwidth]{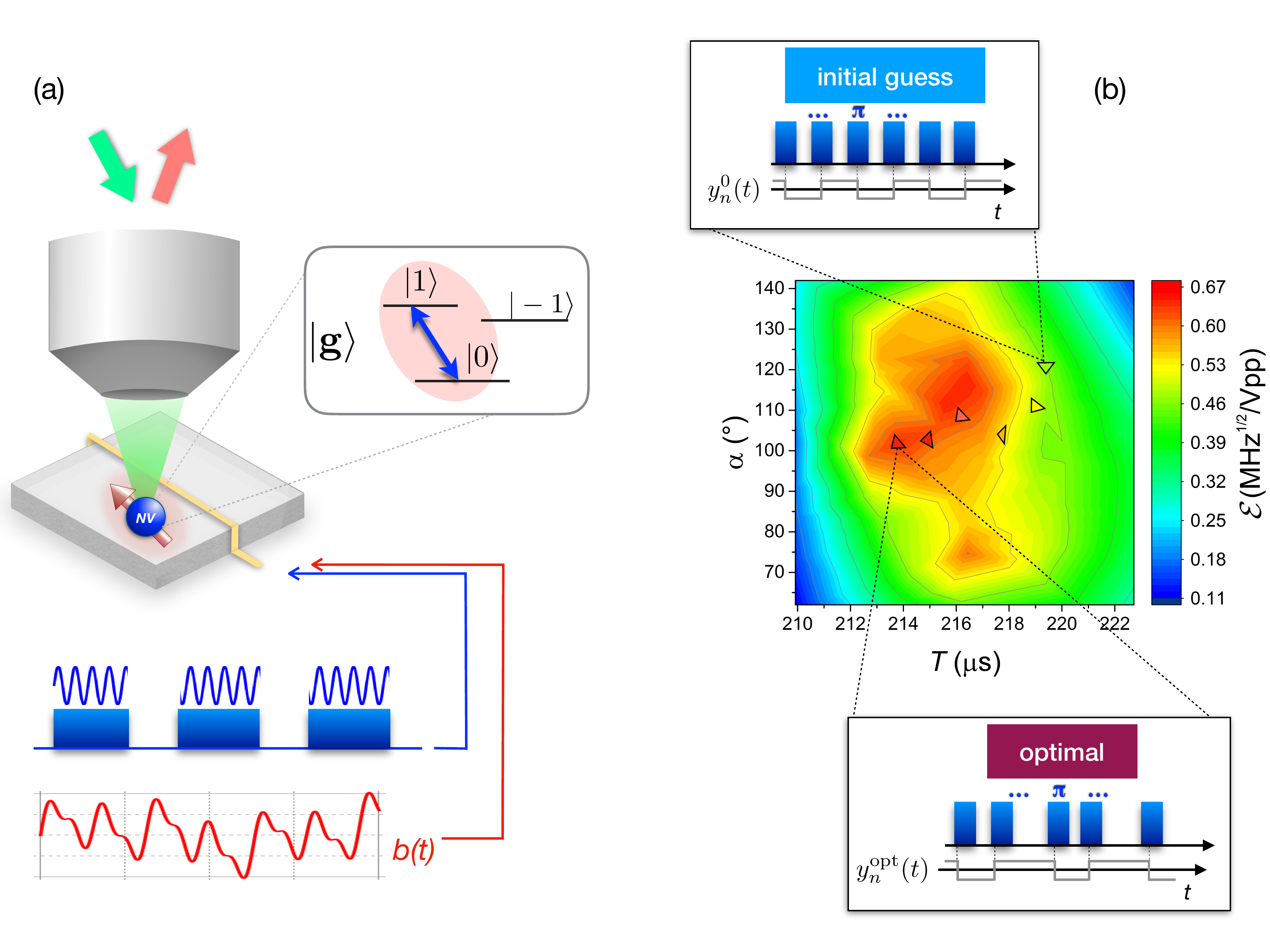}
\end{center}
\caption{{\bf One-qubit optimization strategy.} (a) The electronic spin of a single NV center is optically initialized in the $|0\rangle$ state, and read out at the end of the sensing period, by means of a confocal microscope.  An antenna delivers both the resonant control field (in blue) and the target magnetic field to be measured (in red) in the proximity of the spin qubit. (b) Illustration of the optimization protocol. The starting point is an initial guess for the control sequence described by a modulation function $y_{n}^0(t)$, which depends on a given number of parameters. While in the paper we consider more general cases, for the sake of simplicity the central panel shows a search in a two-parameter space (sensing time $T$ and  phase shift $\alpha$) for a Carr-Purcell (CP) control sequence used to detect a monochromatic AC field $b(t)=b\cos(2\pi\nu_0 t+\alpha)$, with frequency $\nu_0=20.5$~kHz, and unknown amplitude $b$ to be measured. The map represents the experimentally-measured inverse sensitivity $\mathcal{E}=C/\eta$ (see text). The algorithm computes the sensitivity $\eta$ under the initial control sequence, then produces and evaluates a number of other trial points $y_{n}^{(i)}(t)$ and moves in the multidimensional parameter space, until global convergence is reached. The final point, described by $y_n^{\textrm{opt}}$, represents the optimal control sequence. } 
\label{Fig1}
\end{figure*}

\section{Optimal control of a qubit sensor with dephasing noise.} 
We use the electron spin states of a negatively-charged nitrogen-vacancy center (NV) in diamond as a sensing qubit of time-varying magnetic fields in the presence of magnetic noise, which induces dephasing of the sensing qubit. While the NV center forms a spin $S=1$, a static bias magnetic field removes the degeneracy of levels with spin projection $S_z =\pm 1$, and a microwave excitation selectively addresses the $S_z=0\rightarrow-1$ transition, therefore the NV center can be effectively described as a single two-level system\footnote{We neglect the hyperfine coupling of the NV electronic spin to nearby nuclear spins. The experiments are performed  in the presence of a static bias magnetic field of $B=39.4$~mT (see  section~\ref{sec:Methods}) aligned along the symmetry axis of the NV center ($\hat{z}$-axis), where we observe full polarization of  the nuclear spin of the Nitrogen-14 composing the NV center~\cite{Poggiali17}\label{fn:footnoteHyper}}. The electron spin qubit can be prepared in a well defined initial state, coherently manipulated, and read out~\cite{Doherty13}.

We can model the coupling of the spin qubit with a general, time-dependent, external field $b(t)=b\, f(t)$ to be measured, and the coupling with noise, through the Hamiltonian 
\begin{equation}
\mathcal{H} = \gamma\, b(t)\sigma_z + \gamma \,\beta(t)\sigma_z,
\label{Ham}
\end{equation}
where $\beta(t)$ is a stochastic variable with power spectral density $S(\omega)$ in the frequency domain, $\gamma=2.81\times10^{4}$~Hz/$\mu$T is the NV gyromagnetic ratio, and $\sigma_z$ is the $z$ component of the spin operator, $\hat{z}$ being the NV symmetry axis. Performing metrology means reaching a compromise between two conflicting tasks, {\it i.e.}, minimizing the noise effects while maximizing the signal stemming from the field, during the sensing time. Here, in particular, we assume to know the temporal dependence $f(t)$ of the field, and we aim at measuring its amplitude $b$ (we are thus interested in a parameter estimation task). 

While different control strategies can be used for sensing, here we consider control via pulsed dynamical decoupling, which is realized with series of $\pi$-pulses that repeatedly flip the spin, thus reversing its evolution~\footnote{Here we limit the notion of ``optimal'' control to optimality over this restricted choice of control strategies. We are supported in our choice of strategy by the great success that dynamical decoupling has obtained in quantum sensing. While allowing general control fields could lead to a true optimal solution, it would also complicate the search space and its computational cost.  While we do not demonstrate the realization of the ultimate limit of sensitivity, that is, the Quantum Cr\'{a}mer-Rao bound~\cite{Holevo82}, our optimal control method remarkably enhances sensitivity in a number relevant and demanding experimental scenarios.\label{fn:footnoteQCR}}. The control field can be thus described by a modulation function $y_n(t)$, with a sign switch at the position of each $\pi$ pulse, indicating the direction of time evolution, forward or backward. The squared Fourier transform of $y_n(t)$ defines the filter control function $Y_{n,T}(\omega)$.

The phase accumulated by the spin during the sensing time $T$, under the action of the control field is 
\begin{equation}
\varphi_n(T)=\int_0^T \gamma \,b(t) \,y_n(t) dt \equiv b \, \phi_n.
\label{phiaccu}
\end{equation}
To read out the phase due to the target field, we embed the control sequence within a Ramsey interferometer, which enables the mapping of the phase accumulated into observable population of the spin projection $S_z=-1$.

As said, during the sensing process, the sensor qubit is also subject to noise. In the case of the NV center, this is mainly due to the nuclear spin bath that generates a stochastic time-varying field. Therefore, the qubit acquires a random  phase during its coherent evolution, which leads to a reduction of the observed population.

The state of the qubit after the sensing process is described, as in a Ramsey interferometer, by population and coherence of the density matrix 
\begin{equation}
\rho_{1,1}(T)=\frac{1}{2},\;\;\; \rho_{1,2}(T,b)=\frac{-i}{2}e^{-i \varphi_n(T)}e^{-\chi_n(T)},
\label{rho1}
\end{equation}
where $\chi_n(T)$ is temporal coherence function,  describing noise-induced decoherence, that also depends on the control field through $Y_{n,T}(\omega)$
\begin{equation}
\chi_n(T)=\int d\omega\,S(\omega)\,\left|Y_{n,T}(\omega)\right|^2/(\pi \omega^2).
\label{chi}
\end{equation}
Thus, a projective measurement on the $\sigma_x$ basis, $| \pm \rangle=(|0\rangle\pm|1\rangle)/\sqrt{2})$ gives a signal 
\begin{equation}
s(T)=\langle+|\rho(T, b)|+\rangle=\frac{1}{2}\left(1+e^{-\chi_n(T)}\cos\varphi_n(T)\right).
\end{equation}

To assess the quality of  parameter estimation, as achievable under a given control protocol and within the experimental constraints, we can evaluate the Fisher information (FI)~\cite{Braunstein94,Holevo82} associated with the measurement,
\begin{equation}
F_N=\sum_x \frac{1}{p_N(x|b)}\left(\frac{\partial p_N(x|b)}{\partial b}\right)^2.
\label{fisher0}
\end{equation} 
Here $p_N(x|b)= \mbox{Tr}[E_x^{(N)} \rho_b^{\otimes^N}]$ are conditional probabilities of obtaining $x$ as measurement result for a given field $b$ over $N$ repeated measurements, $E_x$ being the measurement estimator and $\rho$ the density matrix of each independent copy of the system. 
Sensitivity, that is, the minimum detectable signal per unit-time, is simply related to the Fisher information by
\begin{equation}
\eta=\frac{\sqrt{\mathbb T}}{ \sqrt{N F_N}},
\label{eta0}
\end{equation} 
where ${\mathbb T}=NT$ is the total sensing experiment time. For the one-qubit sensing schemes we are considering, this reduces to 
\begin{equation}
\eta=\min\left\{\frac{\Delta s}{\partial_b s}\right\}\sqrt{T}=\frac{e^{\chi_n(T)}}{|\phi_n|}\sqrt{T}.
\label{eta}
\end{equation}

This is indeed the cost function that we want to minimize. In practice, for a given field $b(t)$, we are searching for the optimal control field that steers the spin trajectory of the electronic spin on the Bloch sphere in such a way that, while the accumulated phase $\varphi_n(T)$ is maximized, the effect of non-markovian noise described by $\chi_n(T)$ is minimized. 

To this purpose, we have designed a direct and fast search method that looks for the optimal modulation function $y_n^{\textrm{opt}}(t)$ 
that minimizes the cost function $\eta$.
We have investigated various multi-dimensional parameter spaces, up to dimension $M=51$, and analyzed which optimization parameters ({\it e.g.}, total sensing time, $\pi$-pulse positions, signal phase, signal trigger time) do provide the largest improvement without requiring excessive computational resources, as we will detail in the following. The constraints of the parameter space are chosen to describe realistic experimental condition. The search of the optimal control field is performed by means of a Simplex (Nelder Mead) minimization numerical algorithm that allows for reaching global convergence in the parameter space, as illustrated in Fig.~\ref{Fig1}b. The method requires a precise knowledge of the temporal coherence function of the electronic spin sensor, which depends on the noise spectrum induced by its spin bath, as detailed in the Method section~\ref{sec:spectrum}.

\section{Results}\label{sec:results}

\begin{figure*}[t!]
\begin{center}
\includegraphics[width=.9\textwidth]{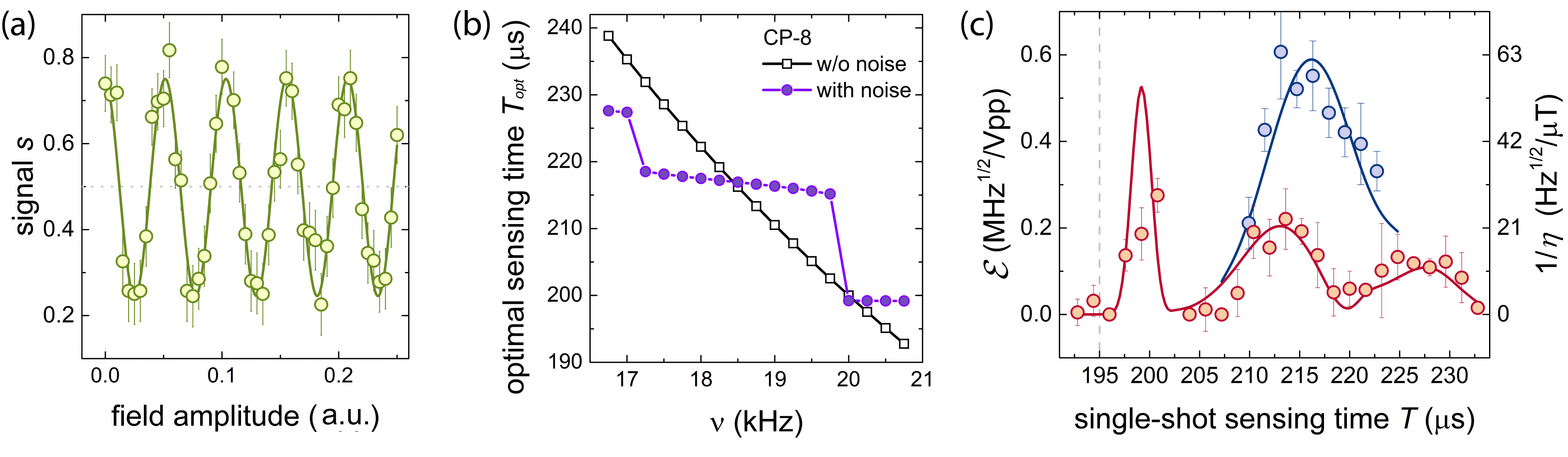}
\end{center}
\caption{{\bf Optimized sensing of monochromatic fields.} (a) Experimental signal measured in the presence of a monochromatic target AC magnetic field, $b(t) = b \cos (2\pi \nu t + \alpha)$, with $\nu=9.24$~kHz and $\alpha=0$, as a function of the target magnetic field amplitude $b$. Here, the spin sensor is controlled with a Carr-Purcell sequence of $n=4$ equidistant $\pi$-pulses. Dots are the experimental data, the curve is a cosinusoidal fit. Error bars are the statistical errors over $2\times 10^5$ repeated measurements. (b)  Theoretical prediction of the optimal sensing time of a Carr-Purcell sequence of $n=8$ equidistant $\pi$-pulses (CP-8), calculated for a target AC magnetic field $b(t) = b \cos (2\pi \nu t + \alpha)$, as a function of the AC frequency $\nu$. The black empty squares are the optimized solutions of the sensing problem when neglecting the presence of the noisy environment, and the black curve represents the expected optimal time, $T_{opt}= n/(2\nu)$, with no fitting parameters. Purple dots are optimized solutions including the noise-induced decoherence of the spin qubit (the line is a guide to the eye). (c) Inverse sensitivity, experimentally measured,  $\mathcal{E}$ (left side vertical axis, see text) and theoretical, $1/\eta$ (right side vertical axis) in the presence of a cosinusoidal field $b(t)$ of frequency $\nu_0=20.5$~kHz under CP control, with $\alpha=0$ (red dots, experiment; red line, theory), and with $\alpha=102^\circ$ (that is, an initial delay time $t_0=0.28/\nu_0$) resulting from optimization (blue dots, experiment; blue line, theory). The experimental error bars come from the  slope uncertainty of the experimental signal $s$.
 } 
\label{Fig2}
\end{figure*}

The experiment is sketched in Fig.~\ref{Fig1}a (see also Method Sec.~\ref{sec:expt}). The electronic spin of a single NV centre is optically initialized and read out by means of a confocal microscope. The spin qubit is coherently controlled with a resonant field, and radiated with the off-resonance time-varying target magnetic field to be measured. 
We obtain sensitivity of the spin qubit to the target field by sweeping the amplitude of the magnetic field and measuring the slope of the signal $s(T)=\left(1+e^{-\chi_n(T)}\cos(\phi_n \,b)\right)/2$ at the points of maximum slope (where $s=0.5$), as shown in Fig.~\ref{Fig2}a. From this quantity we extract the experimental observable $\mathcal{E}=\max\{\partial_b s\}/\sqrt{T}$. 
Following Eq.~\ref{eta}, this experimental observable is simply related to sensitivity through the relation $\mathcal{E}=C/\eta$, where $C$ is the only calibration constant, independent of the target field strength and of the control sequence (see Method Sec.~\ref{sec:sensitivity}).

\subsection{Optimized sensing of an oscillating field \label{sec:AC}}
We first focus on the simple case of monochromatic sinusoidal signals $b(t) = b \cos (2\pi \nu t + \alpha)$, in a rather wide frequency range, $\nu = 20 - 125$~kHz (see Supplemental Material). We start from a common pulsed dynamical decoupling sequence, the Carr-Purcell (CP) multi-pulse sequence, originally devised in nuclear magnetic resonance~\cite{Carr54,Slichter}, which has been demonstrated to extend the qubit's coherence~\cite{Viola98} and has been successfully employed in sensing to measure monochromatic AC magnetic fields (see, {\it e.g.},~\cite{Taylor08, Zhao12,Kolkowitz12,Taminiau12,Loretz14,Shi15}). CP is composed by $n$ $\pi$-pulses, equally spaced by $\tau=T/n$, which periodically flip the spin qubit. This kind of sequence is highly selective in frequency: its filter function $Y_{n,T}(\omega)$  is indeed peaked at $\nu=1/(2\tau)$~\footnote{For a bare CP sequence, the filter function is $Y_{n,T}=[\sin(\pi\nu T)/(\pi\nu T)(1-\sec(\pi \nu T/n)) \cos(\pi\nu T+\alpha)]^2$, which is peaked at $\nu=k/2\tau$, with $k$ being an odd integer index, where the higher harmonics have decreasing weight~\cite{Degen17}\label{fn:footnote2} }\setcounter{fnnumber}{\thefootnote}. In the case of interest, the qubit is subject to colored noise  due to a nuclear spin bath in diamond, where the main component is  due to Carbon-13. If the target signal has frequency close to the center of the noise spectrum, CP control may be not the best choice, since the sequence achieving noise cancellation also leads to a significant attenuation of the signal to be measured.

As a warmup for the full optimization, we optimize the control over a restricted space of two parameters, the sensing time $T$, and the initial phase shift $\alpha$, with a fixed number of pulses, $n=8$. First, fixing $\alpha=0$ we find the optimal sensing time as a function of the AC frequency, as reported in Fig.~\ref{Fig2}b. 
Taking into account decoherence effects appreciably modifies the optimal sensing time (purple curve), compared to the results obtained in the absence of noise sources (black curve), where the optimization routine recovers the expected analytic solutions $T_{opt}= n/(2\nu)$. Then, we optimize both $T$ and $\alpha$. 
To evaluate the global convergence of the optimization, we have also mapped $1/\eta$ in the two-dimensional (2D) parameter space ($T,\alpha$). Figure~\ref{Fig1}b shows this map for an AC field of frequency $\nu_0=20.5$~kHz. This allows the results of the optimization to be compared with the brute-force approach of an extensive search in the parameters space. 
The full optimization of the two parameters, including noise effects, is able to find the global minimum of sensitivity (the optimized parameters are $T=216$~$\mu$s and $\alpha=102^\circ$, corresponding to an initial delay time $t_0=0.28/\nu_0$ of the control sequence).
Figure~\ref{Fig2}c shows some cuts of the previous 2D-map as a function of $T$, with $\alpha=0$ (red solid line), and with $\alpha=102^\circ$ (the optimal value resulting from the numerical search, blue line). 
Using Eq.~\ref{eta}, we also calculate the experimental observable $\mathcal{E}=C/\eta$ (left side vertical axis in Fig.~\ref{Fig2}c), which can be directly compared with the experimental findings at fixed $\alpha=0$ (red dots), and with $\alpha=102^\circ$ (blue dots). We find good agreement of the experiments with the results of the optimization. We also remark that even in the simple case of one parameter optimization, including the noise effects yields a different optimal sensing time than what calculated in the absence of noise ($T= n/(2\nu_0)$, gray vertical dashed line), and this is also reflected in the observed experimental peak of $\mathcal{E}$ - vs - $T$.
 
\subsection{Optimized sensing of multitone AC signals.} \label{sec:AC3}
We then tackle the more complex task of measuring arbitrary time-dependent signals. We consider multitone magnetic fields, in the form $b(t)=b\sum_i^m w_i \cos(2\pi \nu_i t+\alpha_i)$, where $m$ is the number of Fourier components, $b_i=b\,w_i$ their amplitudes (with $\sum_iw_i=1$), $\nu_i$ their frequencies, and $\alpha_i$ the initial phases. 
We employ our optimization tool to engineer optimal control sequences of non-equidistant $\pi$-pulses that may extract information from multitone target signals, while refocusing spin dephasing better than common dynamical decoupling solutions.

\begin{figure*}[t!]
\begin{center}
\includegraphics[width=.8\textwidth]{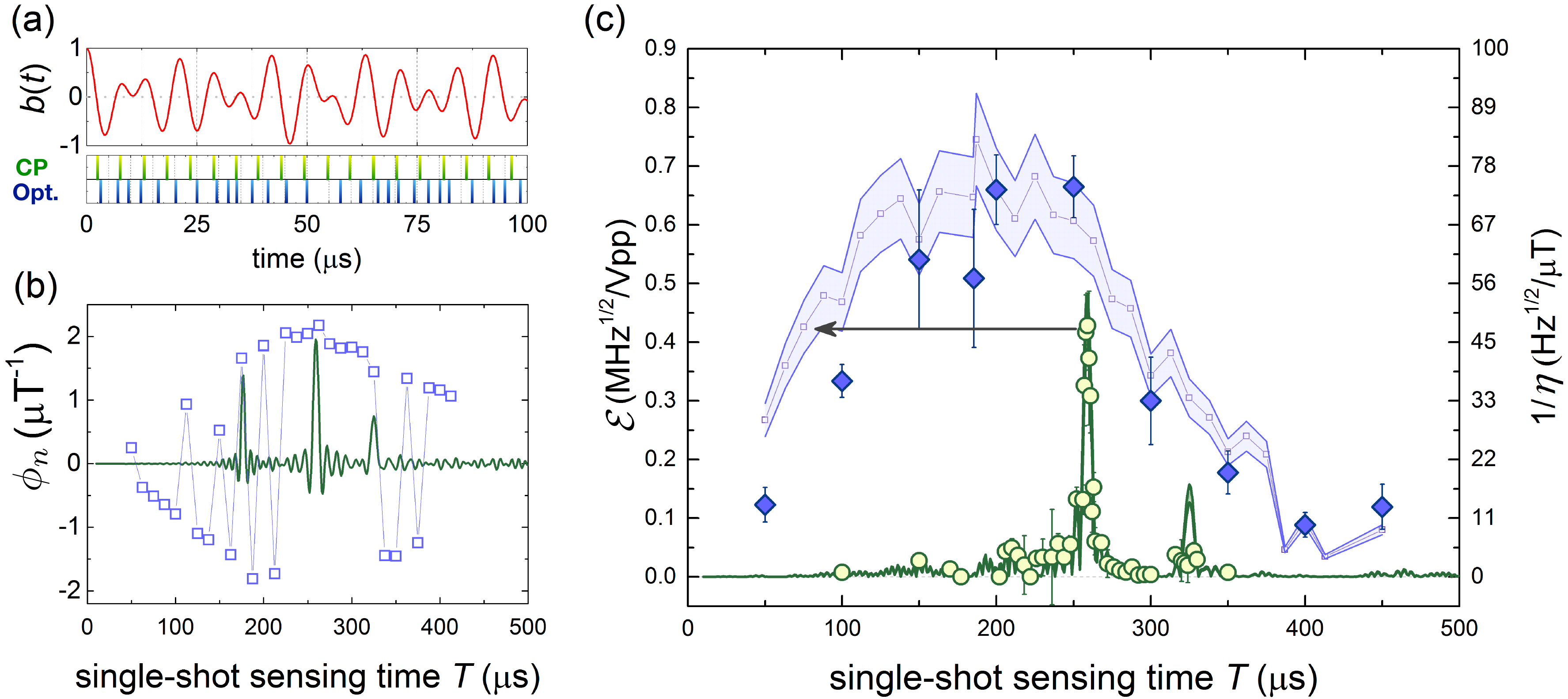}
\end{center}
\caption{{\bf Optimized sensing of multitone AC fields.} (a) Upper panel: Sample multitone target field, $b(t)=b\sum_i^3 w_i \cos(2\pi \nu_i t+\alpha_i)$, with $\alpha_i=0$, frequencies $\nu_i=(77; 96; 141)$ kHz and amplitudes $w_i=(0.45; 0.43; 0.12)$ G, respectively. Bottom panel: in green, position of the first 19 $\pi$-pulses of a Carr-Purcell sequence of 50 equidistant $\pi$-pulses (CP-50) with optimized sensing time ($T=260$~$\mu$s); in blue, position of the first 27 $\pi$-pulses of an optimal control sequence of 50 $\pi$-pulses with optimized time intervals and optimized initial phase  ($T=187$~$\mu$s, $\alpha_i=0.3$). (b) Phase $\phi_n(T)=\varphi_n(T)/b$ accumulated by the spin qubit sensor during the sensing time $T$ in the presence of the field $b(t)$, under a control field of $n=50$ $\pi$-pulses, in the cases of   CP-50 (solid line), and  optimized control  (blue squares). (c) Experimentally measured $\mathcal{E}=C/\eta$ in the presence of the field $b(t)$ under CP (dots) and optimized control (diamonds). The curves represent the theoretical prediction for $\mathcal{E}$, for CP (solid green line) and optimized control (blue line) respectively, obtained by rescaling $1/\eta$ (right-hand side vertical scale) with the unique factor $C$. The shaded area takes into account the experimental uncertainty due to $C$ (see Method section~\ref{sec:sensitivity}).
\label{Fig3}}
\end{figure*}

As said, common multipulse control sequences like CP are in general highly selective in frequency. For this reason, these sequences may exhibit sub-optimal performances when probing a multitone target field, due to attenuation of some frequency components. In addition, increasing the interrogation time to enable a larger phase accumulation, thus improving measurement sensitivity,  also further narrows the width of the filter function $Y_{n,T}$ as $\sim1/(T)$~[\textcolor{blue}{\thefnnumber}]. If the magnetometry task consists in measuring the signal amplitude of a spectrally characterized source as we are considering here, CP collects information about mostly one frequency component at a given sensing time.
When fixing the number of pulses $n$ and sweeping the total time $T=n\tau$, the phase accumulated by the spin qubit sensor under CP control reflects the spectral composition of the signal, showing peaks at times $\tau_i= 1/(2\nu_i)$. This is exemplified in Fig.~\ref{Fig3}b, where we consider  a  field made of $m=3$ Fourier components under a CP  train of $n=50$ pulses (green solid line). However, the sensor's decoherence influences the final sensitivity by suppressing the response of one of the three frequency components, as shown both in theory and in experiment (green solid line and yellow dots in Fig.~\ref{Fig3}c).  

For this kind of scenario, optimal control strategies offer a key advantage. Optimal control can indeed be exploited to find optimal distributions of the $\pi$-pulse positions. Sequences of non-equally distributed pulse spacings, devised by means of analytical models, have been indeed demonstrated to correct for selectivity of CP in certain cases~\cite{Zhao14,Casanova15,Ajoy17}. 
In the case of multitone AC signals to be measured, such sequences enable to simultaneously collect signal from all the various frequency components thus achieving a faster phase accumulation. The results of optimization for the multitone field considered above are shown in Fig.~\ref{Fig3}  \footnote{Different multitone AC fields are considered in the Supplemental Material.}. For the optimization, we keep the number of pulses fixed to $n=50$, and optimize all the $\pi$-pulse positions and the initial phase ($\alpha_i=\alpha$), in all 51 free parameters. 
We impose the time intervals around to each $\pi$-pulse to be symmetric with respect to the pulse position in order to ensure cancellation of static noise and better refocusing of low-frequency noise (see Method Sec. \ref{sec:sensitivity}). 
As shown in Fig.~\ref{Fig3}b, the optimization method leads to a remarkable improvement in the accumulated phase per unit field amplitude $\phi_n$ (blue squares) compared to CP (green line), over an extremely wide range of sensing times. The overall-optimal control sequence (obtained with sensing time $T=187$~$\mu$s, and phase shift $\alpha=0.3$) realizes a sensitivity $\eta_{opt}^{best} = 12$ nT/$\sqrt{\mbox{Hz}}$. 
Since each of our sensing experiments is typically obtained by averaging over $N=2\times10^5$ measurement shots, the optimized control sequence enables the measurement of a local field of $2$~nT. 
The improvement in sensitivity is almost two orders of magnitude compared to sensitivity of CP ($\eta_{CP}=82.6$~mT/$\sqrt{\mbox{Hz}}$) at the same sensing time. 
We remark that the best sensitivity obtained with CP control is still a factor of $1.75$ worse than the best sensitivity achieved with the optimized control ($\eta_{CP}^{best} = 21$ nT/$\sqrt{\mbox{Hz}}$), and with an acquisition time ($T= 260$ $\mu$s) that is 40\% longer than the  optimal sequence. In addition, optimized control is able to achieve the same $\eta_{CP}^{best}$  three times faster then CP ($T= 75$ $\mu$s, compared to $T= 260$ $\mu$s, see black arrow in Fig.~\ref{Fig3}c).  
Thus, optimal control tools have allowed to obtain both a remarkable enhancement of sensitivity and a speed-up of the measurement of multitone AC fields.

\begin{figure*}[t!]
\begin{center}
\includegraphics[width=.8\textwidth]{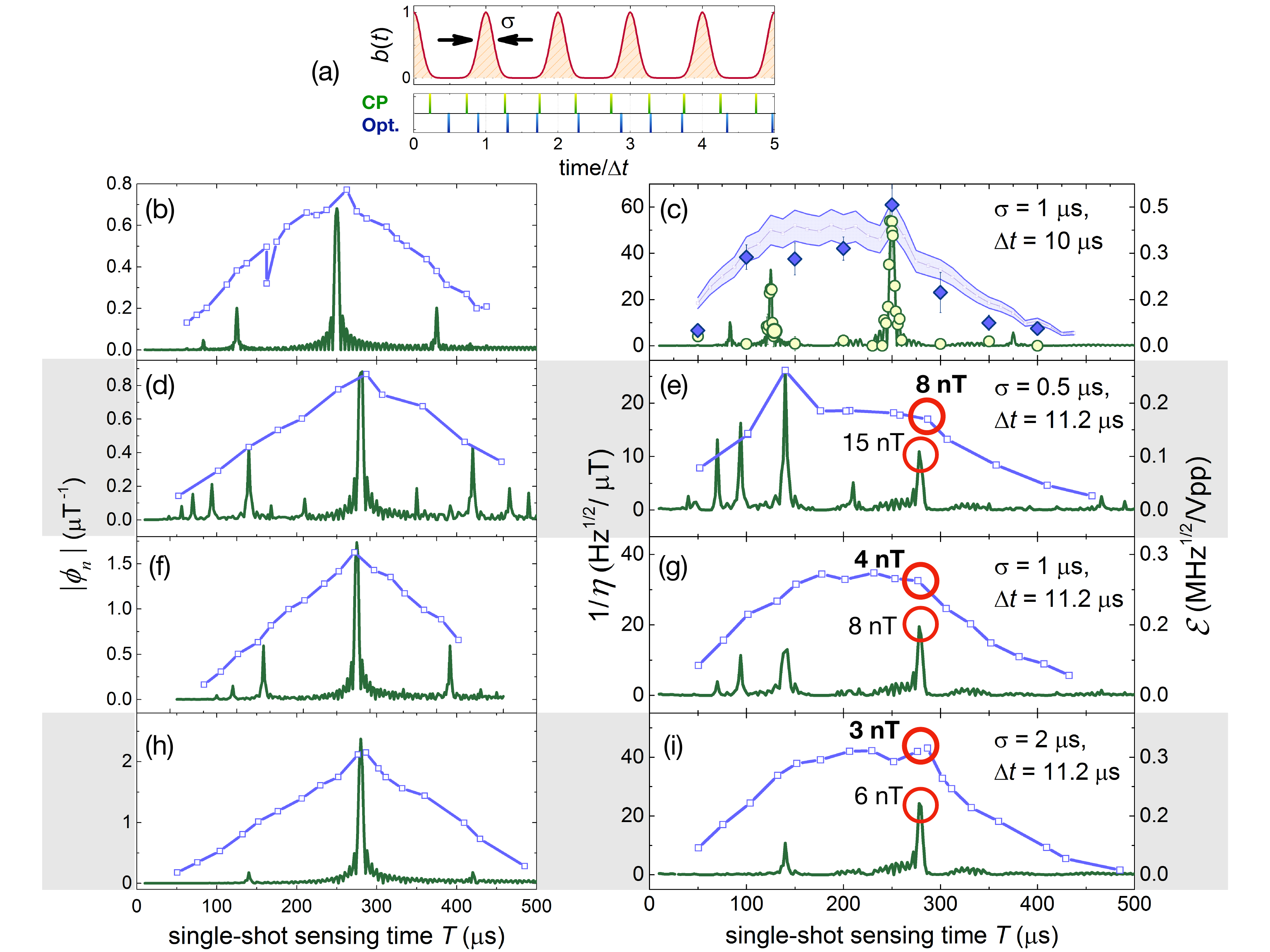}
\end{center}
\caption{{\bf Optimal sensing of gaussian impulses}. (a) Upper panel: Target signal made of a train of gaussian impulses of width $\sigma$ and repetition rate $1/\Delta t $. Bottom panel: position of the first 10 $\pi$-pulses of a CP control sequence of 50 $\pi$-pulses (CP-50), with total sensing time $T=280$~$\mu$s optimized to sense a train of gaussian impulses with $\sigma=2$~$\mu$s, $\Delta t=11.2$~$\mu$s (in green), and optimal position of the first 10 $\pi$-pulses of a control sequence of 50 $\pi$-pulses (in blue), optimized to sense the same target   field (50 optimization parameters). (b)-(i) Modulus of the phase accumulated by the spin qubit sensor in the presence of the target field (left panels), and the inverse of sensitivity 1/$\eta$ (right panels), as a function of the sensing time $T$, under CP-50 control (green solid curves), and under optimized control (blue lines with squares) with bounds $\tau\in(0.6-10)$~$\mu$s (see  Method Sec.~\ref{sec:Methods}). 
In (c), measurement of the experimental observable $\mathcal{E}$ as resulting from CP experiments (yellow dots), and from optimized control (blue diamonds), scaling as indicated on the right-hand side vertical axis. The target field parameters of panels (b-i) are: (b-c), $\sigma = 1.0$~$\mu$s, $\Delta t= 10$~$\mu$s; (d-e), $\sigma = 0.5$~$\mu$s, $\Delta t= 11.2$~$\mu$s;  (f-g), $\sigma = 1.0$~$\mu$s, $\Delta t= 11.2$~$\mu$s;  (h-i), $\sigma = 2.0$~$\mu$s, $\Delta t= 11.2$~$\mu$s. \label{Fig4}}
\end{figure*}

\subsection{Optimized sensing of trains of magnetic impulses.}\label{sec:gauss} 
We have applied optimal control to the different scenario where the target  magnetic field is a train of impulses. This is in general the case of the temporal shape of electric and magnetic fields associated to cardiac, neural, and nervous activities of human and animal organs  \cite{Wikswo80,Bison09,Jensen16,Barry16}. For this kind of application, the NV sensors may offer the remarkable advantages of subcellular spatial resolution, in addition to high sensitivity, and biocompatibility \cite{Schrand09}.
 
As illustrative models for these biological applications, we consider a train of gaussian-shaped impulses. The target field is thus of the general form, shown in Fig.~\ref{Fig4}a,  
\begin{equation}
b(t)= b\sum_{i=0}^{m_{r}} e^{-\frac{(t-i \Delta t)^2}{2\sigma^2}}
\label{multipulse}
\end{equation}
where $1/\Delta t$ is the repetition rate, and $m_r$ is the number of repetitions, with $m_r\,\Delta t\gg T$. In this case, standard dynamical decoupling may be underperforming, since the target signal $b(t)$ is positive-defined in the whole temporal domain, thus the product $y_n(t) b(t)$ may be alternately positive and negative, reducing the accumulation of useful phase (see Eq.~\ref{phiaccu}). In other words, each time a $\pi$-pulse reverses the spin dynamics, it can partially cancel not only the effect of unwanted noise, but also the phase associated to the field to be measured.  

Figure~\ref{Fig4}b-i compares the results of CP control (green-colored curves) and optimal control (in blue), when varying the width $\sigma$ of the target gaussian pulse train and its repetition rate $1/\Delta t$.  In this case as well, we evaluate the effect of control sequences made of $n=50$ $\pi$-pulses. In the optimization, all the time intervals between the control pulses, symmetrized around the $\pi$-pulse positions, are free parameters. Left panels represent the modulus of the phase accumulated by the spin qubit sensor per unit of the target magnetic field amplitude $|\phi_n|$, whereas right panels represent the inverse of sensitivity $1/\eta$, as a function of the sensing time $T$. 

Optimal control outperforms CP in accumulating useful phase due to $b(t)$ over a large sensing time range. Both CP and optimal control do lead to their largest phase accumulation when $T\simeq n\Delta t /2$, where they give similar results in $\phi_n$. This condition corresponds of having couples of $\pi$-pulses  located in each ``empty" time window between two gaussian pulses of the target field, albeit optimal control corrects in a non-trivial way the distribution of $\pi$-pulse positions to minimize $\eta$, as represented in Fig.~\ref{Fig4}a (CP, green vertical bars; optimal control, blue vertical bars). This way, the $\pi$ pulses partially reverse the spin qubit dynamics due to undesired noise, but do not cancel the phase due to the target field $b(t)$. 

We note that, even when the phase accumulated with CP and with optimal control is comparable, optimal control compensates better than CP for decoherence, leading to better overall sensitivity. While here we did not explore this result further, this seems to indicate that numerically optimized sequences might be useful also for other quantum information tasks, such as building a robust memory.
 Fig.~\ref{Fig4}e, g, and i show that optimal control of the spin qubit improves its sensitivity to gaussian multipulse signals up to a factor of two, enabling the measurement of multipulse magnetic fields down to 3~nT.  As shown in Fig.~\ref{Fig4}c,  the measurement of the experimental observable $\mathcal{E}=C/\eta$ confirms the theoretical prediction of sensitivity both for the CP control (yellow dots) and optimal control (blue diamonds).

\section{Discussion}\label{sec:discussion}

Summarizing, we have devised a versatile and robust method of optimal control for quantum metrology with one qubit, and we have applied this optimal control method to the measurement of weak time-varying magnetic fields with a NV spin sensor.  

In practice, we have introduced an unconventional optimization metric, the qubit sensor's sensitivity. The minimization of sensitivity is made by searching the optimal control field that realizes the optimal compromise between useful accumulation of spin phase due to the external field to be measured, and noise refocusing. The developed optimization algorithm offers the advantage of fast convergence and simplicity.  

We have further investigated the robustness of this method for different kinds of real target fields. Optimal control outperforms standard dynamical decoupling in different scenarios, ranging from multicomponent AC target fields in a wide frequency range of the radiofrequency domain, to trains of impulses, which are illustrative examples of the typical shape of electromagnetic field of interest in biology and physiology. 

In the cases investigated, optimal control enables larger phase accumulation over wide sensing-time windows, as well as better cancellation of the effect of external noise on the spin dynamics. Sensitivity of the qubit sensor under optimized control shows an improvement up to a factor of 2, enabling the measurement of pulsed magnetic field down to amplitudes of 3~nT. The comparison of $1/\eta$ with the experimental observable $\mathcal{E}$ demonstrates the reliability of this optimal control method applied to the NV spin sensor. 

Beyond the results obtained in exemplary situations, our novel method is one of the first extensions of optimal control methods to quantum sensing. This rises novel challenges and opportunities, in particular related to the need for new metrics for optimization as well as the challenge to include non-unitary evolution in the numerical optimization. We underline that our optimization method can be extended to larger multidimensional space of parameters, e.g., one can optimize the number of $\pi$-pulses that flip the spin qubit during the sensing time, according to the target signal to be measured. Moreover, while we always considered control sequences given by series of $\pi$ pulses, our scheme can be also generalized to other control strategies of the NV spin qubit. 

Our strategy can be useful for metrology in the face of more and more demanding requirements for the NV spin qubit for applications to sensing of weak time-varying electric and magnetic fields originating, e.g., from individual biological molecules, neuronal networks, nanostructured anti-ferromagnetic or multiferroic materials, and  can be also applied to other physical platforms, such as ultracold atoms or trapped ions. Furthermore, the demonstrated enhanced protection of the spin qubit against noise-induced decoherence, makes optimal control a strategic tool also for building memories in solid-state systems.

\section{Appendix: Methods}\label{sec:Methods}
\subsection{Experimental setup}\label{sec:expt}
The host diamond crystal used in this study is a mono-crystalline electronic-grade sample (Element Six), grown via chemical vapor deposition, with natural 1.1\% abundance of $^{13}$C impurities and $^{14}$N concentration $\ll5$~ppb. All the experiments have been performed on a single negatively charged NV center,  located at $\sim$ 13.5~$\mu$m below the diamond surface.

We exploit a confocal microscope to focus a 532 nm laser beam on the defect and  {collect} the red fluorescence light coming from the diamond. The laser excitation   initializes the NV spin in the m$_S = 0$ state and we perform state readout  after spin manipulation by measuring the fluorescence intensity with a single photon detector.

A permanent NdFeB magnet produces an external static magnetic field, $B =$ 39.4~mT,  aligned along the symmetry axis of the NV center ($\hat{z}$-axis). The field lifts the degeneracy of $m_s = +1$ and $m_s = -1$ energy levels.

Control of the NV spin dynamics is obtained by irradiating the defect with microwave (MW) pulses. We routinely use MW $\pi$ pulses that repeatedly flip the spin, in order to periodically reverse its temporal evolution and refocus the noise effect. The MW pulses trains are applied through a 60~$\mu$m thin copper wire that works as an antenna. We exploit the same wire to deliver time-varying magnetic fields in the radio frequency (RF) range generated by an Arbitrary Waveform Generator (AWG). These RF signals are the target magnetic fields to be measured by the NV. Using two different terminals of the wire we can simultaneously apply both MW and RF fields.

\subsection{Measuring the sensor noise spectrum}\label{sec:spectrum}

Our optimal control strategy depends on the the knowledge of the coherence function $\chi_n(T)$. 
As shown in the main text in Eq.~\ref{chi}, $\chi_n(T)$ depends on both the noise spectrum and the frequency filter $Y_{n,T}(\omega)$ given by the specific sensing sequence. In order to compute the sensitivity $\eta$ (Eq.~\ref{eta}), the optimization algorithm calculates the value of $\chi_n(T)$ for different trial sequences, thus it needs the noise spectral density $S(\omega)$ as an input. 
Various methods to measure the noise spectrum have been suggested in the literature~\cite{Yuge11,Alvarez11,Young12,Bylander11,Faoro04,Almog11,Bar-Gill12}. We followed here the procedure described in~\cite{Yuge11}. 

The filter function  $Y_{n,T}(\omega) = |y_n(\omega,T)|^2$, where $y_n(\omega,T)$ represents the Fourier transform of the modulation function $y_n(t)$, has a simple form for periodic sequences.  
In the limit of large pulse numbers, it can be approximated by a delta function at the angular frequency $\pi/\tau$, where $\tau$ is the pulse spacing. Then, the coherence signal  decays as $s(t) \sim e^{-t/T_2^{\mbox{\tiny{CP}}}(\tau)}$, where $T_2^{\mbox{\tiny{CP}}}(\tau)$ is a coherence time directly related to the noise spectral density via \cite{Yuge11}
\begin{equation}
\frac{1}{T_2^{\mbox{\tiny{CP}}}(\tau)} \simeq \frac{4}{\pi^2} S (\pi/\tau).
\label{eq:Yuge}
\end{equation}
For each pulse spacing time $\tau$ (which sets the noise frequency that we are considering)  we measured the signal decay as a function of the number of pulses, obtaining $T_2^{\mbox{\tiny{CP}}}(\tau)$. 
By varying the time $\tau$ between MW $\pi$-pulses, we can extract the main frequency components of the noise spectral density  using Eq.~\ref{eq:Yuge}. The spectrum was finally obtained by fitting the raw data with a sum of gaussian functions.

\begin{figure}[b]
\includegraphics[width=8.2 cm]{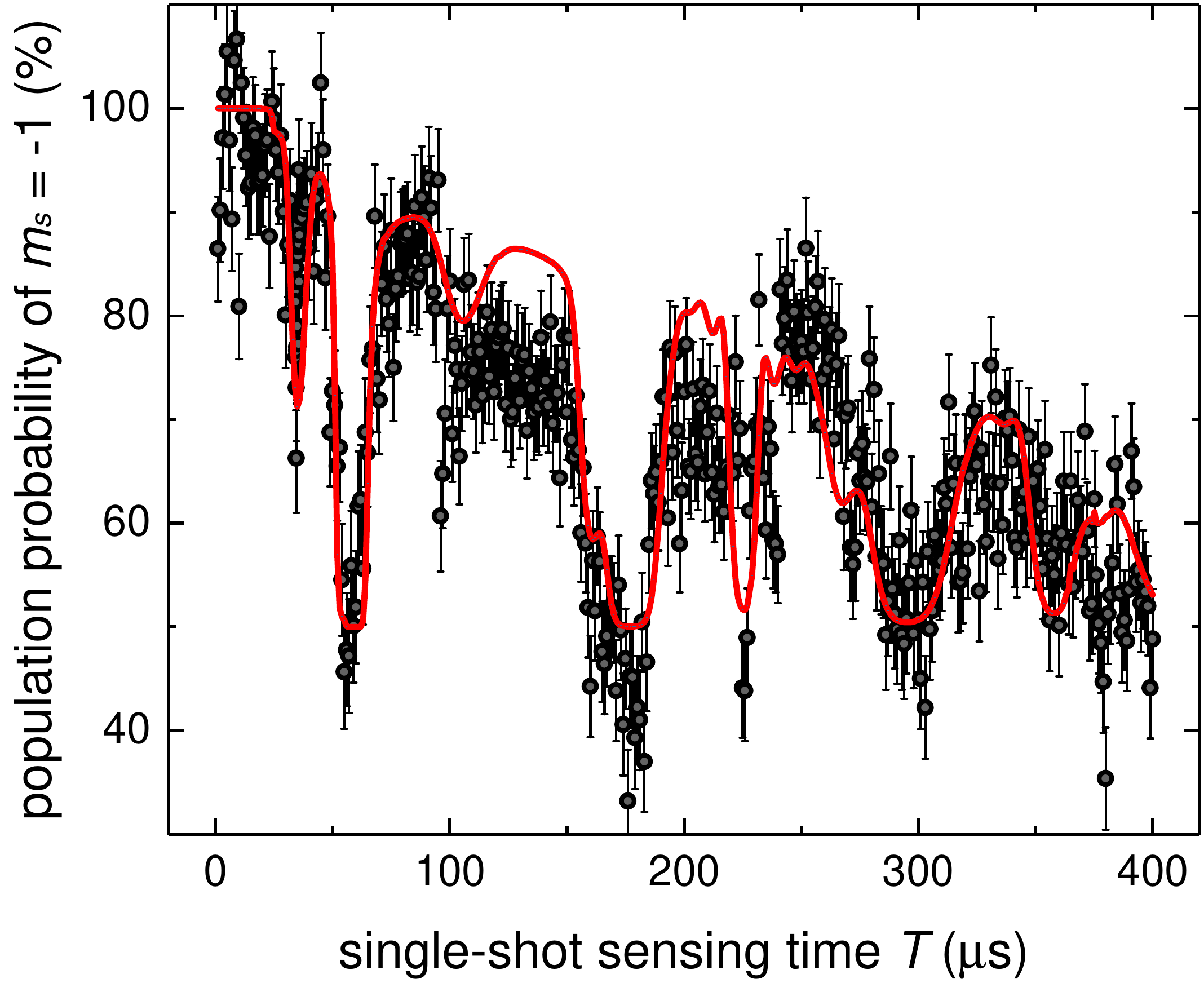}
\caption{{\bf NV spin coherence under CP sequence.}  Probability of the spin projection $m_s = -1$ as a function of the single-shot sensing time, under a CP sequence of $n = 50$ pulses, varying the time $\tau$ between $\pi$-pulses.}
\label{Fig1_meth}
\end{figure}

The experimental spectrum $S(\omega)$ obtained for sequences with different number of pulses showed some variation.
Thus, we further refine the $S(\omega)$ by fitting the decoherence function $\chi_n(T)$ for the CP sequence as in \cite{Reinhard12} (e.g. Fig.~\ref{Fig1_meth}).
This procedure is aimed at correcting $S(\omega)$ for sequence-dependent noise 
and control imperfection, e.g., due to the finite $\pi$-pulses duration~\cite{Loretz15}, which are not taken into account in our model.
We finally find a noise spectrum that, for a fixed number of pulses $n$ is completely independent from the timing at which each $\pi$-pulse occurs, and thus it can be used for sequences very different from CP in the optimization procedure.

We emphasize that good agreement between predicted and measured sensitivity is a further proof of the robustness of our model against imperfection in the empirical function $\chi_n(T)$.

\subsection{Optimization algorithm}\label{sec:algorithm}
The core of our optimal control technique for sensing is an  optimization algorithm that minimizes the sensitivity as a function of the parameters of the control function, e.g. total sensing time, phase of the AC field, and time intervals between $\pi$-pulses.

We use a MATLAB\textregistered\ routine based on the Simplex minimization algorithm to achieve global optimization of the control figure-of-merit, the sensitivity $\eta$. 
The two main ingredients of this quantity are the electron spin phase, $\varphi_n(T)$ and the coherence function $\chi_n(T)$. We consider pulsed control sequences described by the $\pi$-pulse times $\{t_j\}$. For any time-varying external magnetic field $b(t) = b f(t)$ to be measured, we can define $F(t) = \frac{1}{t} \int_0^t f(t')\, dt'$, the integral of the magnetic field (known) temporal profile $f(t)$. 
The phase $\varphi_n=b\,\phi_n$ acquired by the NV qubit can then be calculated for any given control sequence as
\begin{equation}
\phi_n = (-1)^{n+1}F(T)T - 2 \sum_{j=0}^{n+1}(-1)^jF(t_j)t_j
\end{equation}
The coherence $\chi_n(T)$ is instead obtained from the experimentally measured spectrum via Eq.~\ref{chi}. 
From $\chi_n(T)$ and $\phi_n(T)$ we can calculate $\eta$ for each trial sequence according to Eq.~\ref{eta}.

In order to verify the  global convergence of the optimization algorithm, we tested different initial guesses and found the same optimized parameters for a given AC target field. In most cases, we used a constrained search, by setting bounds for each parameter or  constraining the overall result, for instance to keep the total time $T$ constant.

\begin{figure}[h!]
\includegraphics[width=8.2 cm]{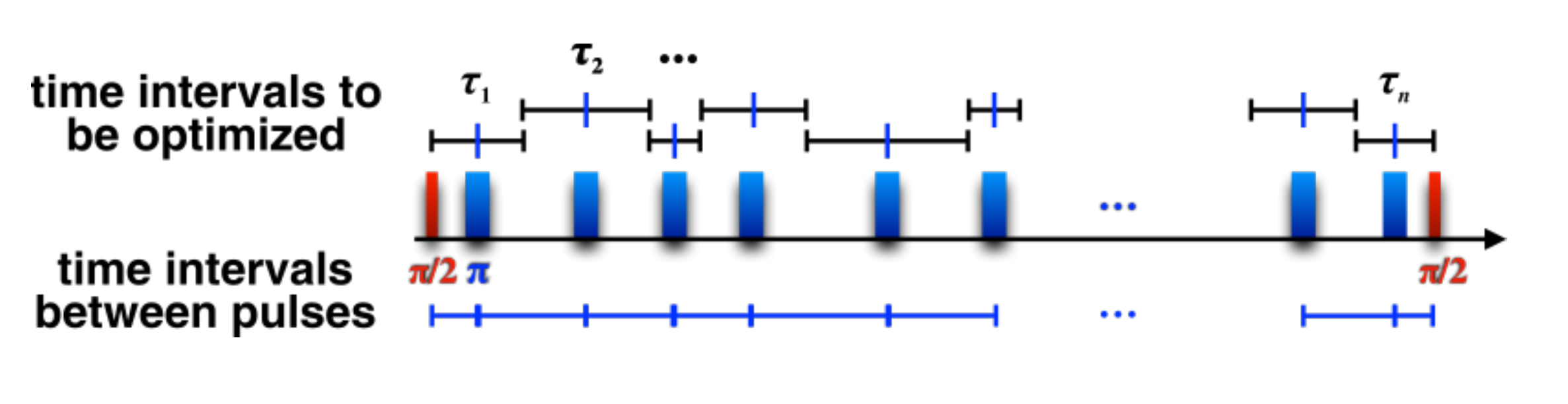}
\caption{{\bf Time intervals $\tau_j$ engineering.} Optimization scheme of the $n$ time intervals $\tau_j$ of a measurement sequence with $n$ $\pi$ pulses. Here, $\tau_j = (t_j+t_{j+1})/2$, where $t_j$ are the $n+1$ time intervals between the $\pi$ pulses, with $j=0,...,n$, and $t_0= t_{n+1} = 0$.}
\label{Fig_seq}
\end{figure}

We note that our procedure is quite general and could be applied to a broad range of sensing scenarios. To demonstrate its reliability, in this work we considered a few exemplary target fields and related control models, varying, e.g., the number of parameters tackled by the optimization algorithm.
We first considered AC fields with a single or multiple frequencies, and we started optimizing $\eta$ as a function of total time $T$ and the AC field phase $\alpha$, while fixing the number of pulses ($n = 8$) and setting $\tau = T/n$ for all  time intervals between the $\pi$-pulses. We then proceeded to allow more flexibility in the optimization, by varying the duration of each time interval between $\pi$-pulses, starting from an initial guess given by a periodic (CP) sequence with $n = 50$. We optimized the time intervals by keeping the time symmetric around each pulse, as shown in Fig.~\ref{Fig_seq}. 
Including also the optimization of the phase of the multitone field, or equivalently, the initial time of the measurement sequence, this optimization manages 51 free parameters. We performed different optimization runs as a function of the total measurement time, keeping $T$ constant in each of them. The only additional constraint that we imposed was to force the times $\tau_j$ between different $\pi$-pulses to be longer than about $10$ times the $\pi$ pulse duration, which in our case means $\tau_j > 600$~ns. This restriction was to ensure that no MW pulse would be very close to each other, as that would have resulted in the $\pi$ pulse to cancel each other giving an effective sequence with a different $n$.

\subsection{Comparison between optimal control theory and experiment}\label{sec:sensitivity}
To experimentally validate the optimal control, we compared the optimized sensitivity $\eta$ with the corresponding measured quantity. However, since we do not have an independent measure of the local amplitude of the magnetic field at the position of the NV center, in the experiment we measure $\mathcal{E} = \mbox{max}\{\partial_b s\}/\sqrt{T} = C/\eta$, where $s$ is the normalized signal and $C$ represents a conversion factor between the generated RF field amplitude and the unknown magnetic field at the defect. 
As $C$ does not depend on the control sequence, it can be evaluated once and then used for all the control scenarios  considered in the paper.

In particular, we estimated $C$ from the experimental results for CP sequences. We evaluated experimentally $\mathcal E(T)$ as a function of the sequence total time $T$ and fitted the curve to extract the maximum $\mathcal{E}_{CP_M}$. Similarly, we evaluated the theoretical value of $\eta$ and obtained its minimum $\eta_{CP_M}$. We then defined $C$ as the product $C=\mathcal{E}_{CP_M}\,\eta_{CP_M}$. This procedure allows us to define not only $C$ but also to estimate its uncertainty $\Delta C$, from the fit error. 
We can then compare the (inverse) experimental sensitivity $\mathcal E_i$ and the theoretical sensitivity $\eta_i$ for each control sequence by rescaling the theoretical sensitivity by $C$.

Finally we investigated the effect of the finite MW pulse duration. Considering the case of a gaussian-shaped  train of magnetic impulses under a CP control sequence, we calculated phase accumulated and sensitivity when excluding from the spin evolution the time intervals where the $\pi$-pulses occurs, finding the same theoretical values of $\eta$, for the all considered total sensing times $T$. We underline that this procedure does not correct the model for the contribution to $\chi_n(T)$ given by MW pulses imperfections or finite duration~\cite{Loretz15}, but confirms that the pulses can be considered instantaneous in our picture for the accumulated spin phase $\varphi_n(T)$; we verified that this approximation is valid up to $\pi_{MW} \le 0.5$~$\mu$s for $n = 50$ and $T \sim 250$~$\mu$s and was very effective for pulses duration $\pi_{MW} = 56$~ns and $T \ge 50$~$\mu$s.

\acknowledgements
We especially thank M. Inguscio for enthusiastic support, S. H\'{e}rnadez G\'{o}mez for critical reading, and all the LENS Quantum Gases group for useful discussions. This work was supported by EU-FP7 ERC Starting Q-SEnS2 (Grant n. 337135).

\bibliographystyle{apsrev4-1}
\bibliography{Biblio}
\end{document}